\documentclass[letterpaper, 10 pt, conference]{ieeeconf}  

\IEEEoverridecommandlockouts                              

\usepackage{graphics} 
\usepackage{epsfig} 
\usepackage{times} 
\usepackage{amsmath,amsthm} 

\usepackage{amssymb}  
\usepackage{algorithm}
\usepackage[noend]{algpseudocode}
\usepackage{subcaption}
\usepackage{amssymb}
\usepackage{multirow}
\usepackage[colorlinks=true,linkcolor=black,citecolor=black,urlcolor=blue]{hyperref}

\newtheorem{theorem}{Theorem}
\theoremstyle{definition}

\newtheorem{remark}{Remark}

\newcommand{\f}{{\mathsf{h}}}
\newcommand{\dd}{{\mathsf{d}}}
\newcommand{\e}{{\mathsf{e}}}

\title{RRT* Based Optimal Trajectory Generation with Linear Temporal Logic Specifications under Kinodynamic Constraints
}
\usepackage{graphicx}
\usepackage{svg}
\author{Saksham Gautam$^{\dagger}$, Ratnangshu Das$^{\dagger}$, and Pushpak Jagtap
\thanks{$^{\dagger}$ Authors contributed equally.}
\thanks{*This work was supported in part by ARTPARK. The work of R. Das was supported by the Prime Minister’s Research Fellowship from the Ministry
of Education, Government of India}
\thanks{S. Gautam, R. Das and P. Jagtap are with the Centre for Cyber-Physical Systems, IISc, Bangalore, India {\tt\small \{sakshamg,ratnangshud,pushpak\}@iisc.ac.in}}%
}
\begin{document}

\maketitle
\thispagestyle{empty}
\pagestyle{empty}

\begin{abstract}
In this paper, we present a novel RRT*-based strategy for generating kinodynamically feasible paths that satisfy temporal logic specifications. Our approach integrates a robustness metric for Linear Temporal Logics (LTL) with the system's motion constraints, ensuring that the resulting trajectories are both optimal and executable. We introduce a cost function that recursively computes the robustness of temporal logic specifications while penalizing time and control effort, striking a balance between path feasibility and logical correctness. We validate our approach with simulations and real-world experiments in complex environments, demonstrating its effectiveness in producing robust and practical motion plans. This work represents a significant step towards expanding the applicability of motion planning algorithms to more complex, real-world scenarios.
\end{abstract}


\section{INTRODUCTION}
Recent developments in motion planning have increasingly focused on handling intricate objectives and constraints using advanced formalisms, including temporal logic \cite{baier2008principles, clarke1986automatic}. Due to their expressive nature and precise semantics, temporal logics, such as Linear Temporal Logic (LTL) \cite{pnueli1977temporal}, Metric Temporal Logic (MTL) \cite{koymans1990specifying}, and Signal Temporal Logic (STL) \cite{maler2004monitoring}, have become vital tools in defining desirable behaviors for dynamic systems. The robust theoretical foundations and practical tools available for temporal logic have facilitated its widespread use in motion planning for robotics and control systems.

Current approaches for generating trajectories that satisfy temporal logic specifications include symbolic control techniques \cite{tabuada2009verification}, sampling-based methods \cite{karaman2012sampling, vasile2013sampling}, graph search techniques \cite{khalidi2020t}, and optimization-based methods \cite{wolff2014optimization}. However, although symbolic control techniques can address input constraints, they typically rely on state space abstraction, which significantly increases computational complexity, especially as the dimensionality of the state space grows. Sampling-based techniques, for instance, abstract continuous state spaces into graph structures that are then used to search for feasible paths. However, they disregard the system dynamics and as these graphs become more complex, scalability issues arise, rendering these methods less effective for complicated planning problems. On the other hand, mixed-integer linear programming (MILP)-based techniques \cite{karaman2011linear} encode LTL specifications directly into optimization problems, yielding optimal solutions. Despite their precision, these methods are often limited to specific subclasses of LTL formulas and heavily rely on user-defined parameters, which constrains their applicability in more general scenarios.

\begin{figure}[t]
\centering
\includegraphics[width=0.4\textwidth]{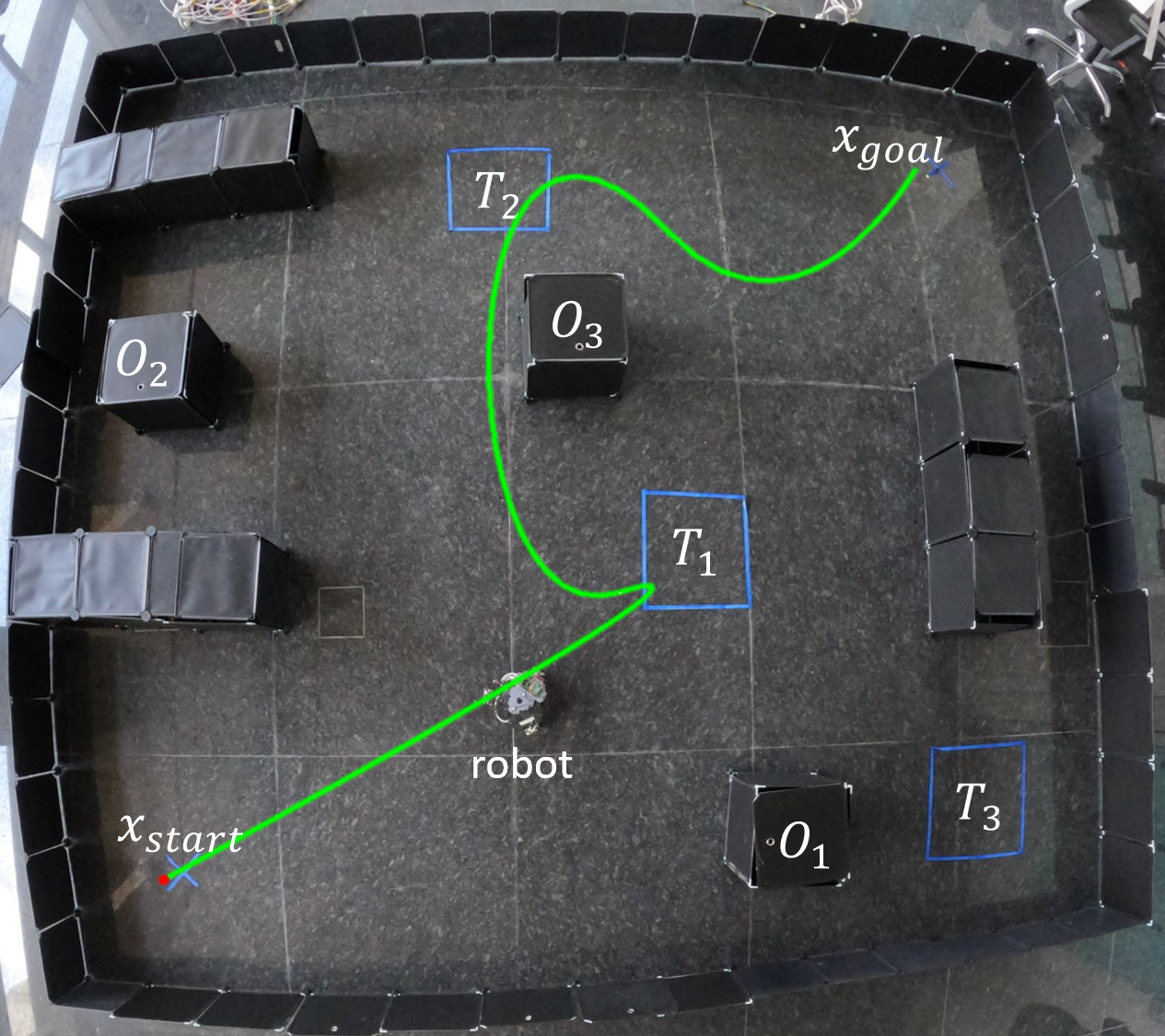}
\caption{Real-world demonstration of optimal, kinodynamically feasible trajectories, meeting LTL specifications.}
\label{fig:rw}
\end{figure}

In recent years, tree-based algorithms have gained attention for incorporating temporal logic preferences into planning, leveraging spatial robustness as a key part of their cost function \cite{karlsson2020sampling, barbosa2019guiding, vasile2017sampling}. Robustness in this context refers to a quantitative measure of how well a trajectory satisfies or violates the given temporal logic specification. It provides a real-valued score that indicates the degree of satisfaction, thus enabling planners to prioritize paths that best meet the desired temporal properties. For example, Karlsson et al. \cite{karlsson2020sampling} formulated a cost function that balances STL spatial robustness with trajectory duration, while others like Vasile et al. \cite{vasile2017sampling} aim to maximize the spatial robustness of STL specifications during motion planning.

While these methods enhance the robustness of trajectory planning, they often do not account for system dynamics, state, and input constraints, which is essential for ensuring the feasibility of the generated paths. A trajectory that satisfies temporal logic specifications but fails to adhere to the system’s kinodynamic constraints may not be implementable in real-world scenarios. Kinodynamic RRT \cite{webb2012kinodynamic} addresses this issue by extending traditional Rapidly-exploring Random Trees (RRT) to consider system dynamics, generating trajectories that not only avoid obstacles but also comply with the system’s motion capabilities.

In this work, we build on these principles by proposing a novel RRT*-based strategy to generate kinodynamically feasible paths that also satisfy temporal logic specifications. Unlike traditional methods, our approach not only focuses on meeting complex temporal logic requirements, through LTL, but also ensures that the trajectories align with the system's physical constraints. This integration of temporal logic and kinodynamic considerations results in a more robust and practical motion planning solution, especially suited for dynamic and constraint-heavy environments.

\section{Preliminaries and Problem Formulation}
\subsection{Notation}
The symbols $\mathbb{N}$, $\mathbb{R}$, $\mathbb{R}^+$, and $\mathbb{R}_0^+ $ denote the set of natural, real, positive real, and nonnegative real numbers, respectively. 
A vector space of real matrices with $n$ rows and $m$ columns is denoted by $\mathbb{R}^{n \times m}$. A column vector with $n$ rows is represented by $\mathbb{R}^{n}$.
The Euclidean norm is represented using $\lVert \cdot \rVert$. For $a, b \in \mathbb{N}$ with $a \leq b$, the closed interval in $\mathbb{N}$ is denoted as $[a; b]$. 
For $a, b \in \mathbb{R}$ with $a < b$, the open interval in $\mathbb{R}$ is denoted as $(a, b)$. 
A vector $x \in \mathbb{R}^{n}$ with entries $x_1, \ldots, x_n$ is represented as $[x_1, \ldots, x_n]^\top$, where $x_i \in \mathbb{R}$ denotes the $i$-th element of vector $x\in\mathbb{R}^n$ and $i \in [1;n]$.

\subsection{System Definition}

Consider the following nonlinear system:
\begin{align}
    \mathcal{S}: \dot{x} = f(x, u), \label{eqn:sysdyn}
\end{align}
where $x(t) \in X$ and $u(t) \in U$ are the system state and input at time $t\in\mathbb{R}_0^+$, respectively, with $X \subset \mathbb{R}^n$ and $U \subset \mathbb{R}^m$ as the bounded state space and input space of the system. 
The bounded workspace $X$ contains obstacles $X_\mathrm{obs}\subset X$ , and the free state is denoted by $X_\mathrm{free} = X \setminus X_\mathrm{obs}$. 

Let $\mathsf{x}_\mathrm{start}$ and $\mathsf{x}_\mathrm{goal}$ be the start and end position of the system trajectory. The trajectory $\pi$ is composed of finitely many segments, each defined by the tuple: $\pi_i= ({x}_i(\cdot),{u}_i(\cdot),\tau_i)$, where $\tau_i$ is the arrival time at the end of segment $i$. The control input along the $i$-th segment is $u_i : [\tau_{i-1},\tau_i] \rightarrow U$ and the corresponding states are $x_i : [\tau_{i-1},\tau_i] \rightarrow X$. If the trajectory is partitioned into $N_s$ segments (i.e., $\pi:=(\pi_1,\pi_2,\ldots,\pi_{N_s})$), then we obtain a sequence of waypoints $(\mathsf{x}_0, \mathsf{x}_1, \ldots, \mathsf{x}_{Ns})$, where, $\mathsf{x}_0 = x_1({\tau_0}) = \mathsf{x}_\mathrm{start}$ and $\mathsf{x}_{N_s} = x_{N_s}(\tau_{N_s}) = \mathsf{x}_\mathrm{goal}$, with $\tau_0 = 0$ as the start time and $\tau_{N_s} = \tau$ as the total trajectory duration.

\subsection{Linear Temporal Logic (LTL)}
Linear temporal logic (LTL) is a high-level language used to describe the desired behavior of a dynamical system. It provides a structured and formal way to express temporal properties, where formulas are inductively built from a set of atomic propositions and combined through Boolean and temporal operators \cite{baier2008principles}. 

In this work, we adopt the Truncated Linear Temporal Logic (TLTL), proposed in \cite{TLTL}, and extend it for continuous trajectories. Unlike traditional LTL, which deals with infinite time horizons, TLTL focuses on finite temporal structures, making it well-suited for encoding domain knowledge for dynamical systems. 
A TLTL formula is defined over predicates of the form $p:= \f(s) < \mu$, where $\f: \mathbb{R}^n \rightarrow \mathbb{R}$ is a function and $\mu \in \mathbb{R}$ is a constant. A TLTL specification is constructed by combining these predicates using logical and temporal operators as follows:
\begin{align}\label{eq:ltlSpec}
\psi ::= &\text{true} \mid p \mid \neg \psi \mid \psi_1  \wedge \psi_2 \mid \psi_1  \vee \psi_2 \mid \Box \psi \mid \Diamond \psi,
\end{align}
where $\neg$ (negation/not), $ \wedge$ (conjunction/and), and $ \vee$ (disjunction/or) are Boolean connectives; and $\Diamond$ (eventually) and $\Box$ (always) are temporal operators. 

Let $x_t := x(t) \in X$ be the state at time $t$ and $x_{[0,\tau]}$ be the state trajectory from time $0$ to $\tau$. The semantics of continuous TLTL are defined as follows:
\begin{align*} &x_{t} \models \f(s) < \mu && \iff \ \f(x_{t}) < \mu,\\
&x_{t} \models \neg\psi && \iff \ \neg(x_{t} \models \psi),\\
&x_{t} \models \psi_1 \wedge \psi_2 && \iff \ (x_{t} \models \psi_1) \wedge (x_{t} \models \psi_2),\\
&x_{t} \models \psi_1 \vee \psi_2 && \iff \ (x_{t} \models \psi_1) \vee (x_{t} \models \psi_2),\\
&x_{[0, \tau]} \models \square \psi && \iff \ \forall t^{\prime} \in [0, \tau] \ x_{t^{\prime}} \models \psi,\\
&x_{[0, \tau]} \models \Diamond \psi && \iff \ \exists t^{\prime}\in[0, \tau] \ x_{t^{\prime}} \models \psi.
\end{align*}

\subsection{Problem Formulation}
Given a nonlinear system $\mathcal{S}$ in \eqref{eqn:sysdyn} with known dynamics and a TLTL specification $\psi$, find a collision-free optimal and feasible trajectory $\pi^*_\mathrm{free}$ and the corresponding set of control inputs between $\mathsf{x}_\mathrm{start}$ and $\mathsf{x}_\mathrm{goal}$ that satisfy $\psi$. 
\begin{align}\label{eq:pi_free}
\pi^*_\mathrm{free} = &\arg\min_{\pi} c[\pi , \psi] \
\text{ s.t., } \nonumber \\
&x(0) = \mathsf{x}_\mathrm{start}, \nonumber \ x(\tau^*) = \mathsf{x}_\mathrm{goal}, \nonumber \\
&\forall t \in [0, \tau^*], \left(x(t) \models \psi  \wedge x(t) \in X_\mathrm{free}  \wedge u(t) \in U \right),
\end{align}
where $\tau^*$ is optimal time of reaching $\mathsf{x}_\mathrm{goal}$, and $c[\pi , \psi] $ is the cost associated with the trajectory $\pi$ and the LTL specification $\psi$. The cost function $c[\pi, \psi]$ has been defined in the subsequent sections. 

\section{Cost Function}
In this section, we discuss the methodology for computing the cost function associated with satisfying a TLTL specification $\psi$ while optimizing the control effort and the time duration of the trajectory. The overall cost function consists of two main components: (1) a control cost $c_{C}[\pi]$ that penalizes the trajectory duration and control effort, and (2) the TLTL cost $c_{TLTL}[\pi, \psi]$ derived from robustness semantics, which quantifies how well a given trajectory $\pi$ satisfies the TLTL formula $\psi$. We then compute the optimal trajectory $\pi^*$ by taking a weighted sum of these two costs.

\subsection{Control Cost}
The first component of the cost function, the control cost $c_{C}[\pi]$, penalizes the trajectory $\pi$ based on its duration $\tau$ and the control effort $u(.)$ required to follow the path, and is defined as \cite{webb2012kinodynamic}:
 \begin{equation}\label{eq:cost_kd}
     c_{C}[\pi]= \int_{0}^{\tau} \left(1 + \frac{1}{2} u(t)^TRu(t) \right) \dd t, 
 \end{equation}
where $R\in \mathbb{R}^{m\times m}$ is a positive-definite matrix. This ensures that the control efforts remain feasible, and the state transitions are smooth.

\subsection{TLTL Cost Computation through Robustness Semantics}
The satisfaction of a TLTL specification can be evaluated using robustness semantics, which measures the degree to which a trajectory satisfies a given temporal logic formula. 

Let the RRT tree be represented by $T$, and each node in this tree by $x_i = x(\tau_i) \in X$, where $\tau_i$ is the time of arrival at node $i$. The parent of a node $x_i$ is denoted as $\mathrm{parent}(x_i) = x(\tau_{i-1})$. To efficiently compute the robustness value along a trajectory $\pi$, we adopt a recursive approach similar to that proposed in \cite{linard2023real}. 
For each node $x_i$ in the tree $T$, we denote the robustness value of the trajectory till the parent node by $ \rho^{\psi}_{\mathrm{parent}}(x_i)$ and recursively define the robustness value for node $x_i$, $ \rho^{\psi} (x_i)$, as:
\begin{align*}
& \rho^{\mathrm{true}} (x_i) &&= \rho_{\mathrm{max}}, \\
& \rho^{\f(x_i) < \mu} (x_i) &&= \mu - \f (x_i), \\
& \rho^{\neg\psi} (x_i) &&= -  \rho^{\psi} (x_i), \\
& \rho^{\psi_1 \vee \psi_2} (x_i) &&= \max( \rho^{\psi_1} (x_i),  \rho^{\psi_2} (x_i)),\\
& \rho^{\psi_1 \wedge \psi_2} (x_i) &&= \min( \rho^{\psi_1} (x_i),  \rho^{\psi_2} (x_i)),\\
& \rho^{\Diamond \psi} (x_{[0, \tau_i]}) &&=\max ( \rho^{\Diamond\psi}_{\mathrm{parent}}(x_i), \max_{t \in [\tau_{i-1}, \tau_{i}]} \rho^{\Diamond \psi}(x(t))),\\
& \rho^{\Box \psi} (x_{[0, \tau_i]}) &&=\min ( \rho^{\Box\psi}_{\mathrm{parent}}(x_i), \min_{t \in [\tau_{i-1}, \tau_{i}]} \rho^{\Box \psi}(x(t))).
\end{align*}

We define the capped robustness $\hat{\rho}^{\psi}(x) = \min( {\rho^{\psi}}(x),0)$, which captures only the non-positive values of the robustness measure. The TLTL cost $c_{TLTL}[\pi, \psi]$ is then defined as the negative of this capped robustness over the trajectory $\pi$:
\begin{equation}
    c_{TLTL}[\pi , \psi]= - \int_{x \in \pi} \frac{\partial}{\partial x} \hat{\rho}^{\psi}(x) \dd x. 
\end{equation}

\subsection{Overall Cost Function}
To compute the optimal trajectory that satisfies the TLTL specification while minimizing control effort and time, we combine the control cost $c_{C}[\pi]$ and the TLTL cost $c_{TLTL}[\pi, \psi]$ into a single objective function $c$. The overall cost function is a weighted sum of these two components: 
\begin{equation}\label{eq:cost_fn}
c[\pi , \psi]= \mu_1 c_{C}[\pi] + \mu_2 c_{TLTL}[\pi , \psi], 
\end{equation}
where $\mu_1 \in \mathbb{R^+} $ and $\mu_2 \in \mathbb{R^+}$ are weighting coefficients that determine the relative importance of satisfying the TLTL specification versus minimizing the control cost.

Minimizing this cost enables the system to plan an optimal path by generating a sequence of waypoints: $(\mathsf{x}_{\text{start}}, \mathsf{x}_1, \mathsf{x}_2, \ldots, \mathsf{x}_{\text{goal}})$, within the free configuration space $X_{free}$. The cost function guides this process, balancing the satisfaction of temporal logic constraints with minimizing control efforts and ensuring smooth transitions.
In the following section, we derive the optimal trajectory and the corresponding control policy, that minimizes this cost while adhering to state, input, and kinodynamic constraints. 

\section{Optimal Trajectory}
Given two successive states $\mathsf{x}_{i-1}$ and $\mathsf{x}_{i}, i = \{1,2,\ldots\}$, the next step is to connect the pair optimally while adhering to kinodynamic constraints. Subsequently, we will prove that taking the union of trajectory segments for each pair of states yields an overall optimal and feasible path for the system.


\subsection{Linear System Representation}
To simplify the nonlinear system dynamics in \eqref{eqn:sysdyn}, we use a first-order Taylor expansion around a nominal operating point $(\hat{x}, \hat{u})$ and approximate them with a linear form as
\begin{equation}\label{eq:sysdynlinear}
    \dot{x}(t) = Ax(t) + Bu(t) + d,
\end{equation}
where $A = \frac{\partial f}{\partial x}({\hat{x}},{\hat{u}}) \in \mathbb{R}^{n\times n}$ is the system matrix, $B = \frac{\partial f}{\partial u}({\hat{x}},{\hat{u}}) \in \mathbb{R}^{n \times m}$ is the control input matrix, and $d = f({\hat{x}},{\hat{u}}) - A{\hat{x}} - B{\hat{u}} \in \mathbb{R}^n$ represents the residual terms.

We now focus on the subproblem of finding the optimal trajectory $\pi^*[\mathsf{x}_{0}, \mathsf{x}_{1}]$ and the associated control $u(t)$, between a pair of successive states, $\mathsf{x}_{0}$ and $\mathsf{x}_{1}$. For simplicity, we consider $\mathsf{x}_{0}$ and $\mathsf{x}_{1}$ in the next 3 subsections. However, this approach can be generalized for any pair of successive states $\mathsf{x}_{i-1}$ and $\mathsf{x}_{i}, i = \{1,2,\ldots\}$.

We proceed through a series of three steps: first, finding the optimal control for a fixed final state and fixed final time, then determining the optimal time required to reach the final state, and finally, calculating the optimal and kinodynamically feasible trajectory between the initial and final states.

\subsection{Optimal Control for Fixed Final State and Time}
Given a fixed arrival time $\tau_1$ and two states $\mathsf{x}_{0}$ and $\mathsf{x}_{1}$, our goal is to determine the optimal control input $u(t)$ and the optimal trajectory $\pi = (x(\cdot), u(\cdot), \tau_1)$, such that the system starts at $x(0) = \mathsf{x}_{0}$, reaches $x(\tau_1) = \mathsf{x}_{1}$, and follows the dynamics $\dot{x} = Ax(t) + Bu(t) + d$. This is known as the fixed final state, fixed final time optimal control problem \cite{lewis2012optimal}.
\begin{theorem}\label{Thm1}
    Given initial state $x(0) = \mathsf{x}_{0}$ and fixed final state $\mathsf{x}_{1}$ at a specified time $\tau_1$, i.e., $x(\tau_1) = \mathsf{x}_{1}$, the optimal control policy $u(t)$ is given by:
    \begin{equation}\label{eq:ut}
      u(t) = \frac{1}{\mu_1} R^{-1} B^T \e^{{A^T(\tau_1-t)}} G^{-1}(0, \tau_1) \left( \mathsf{x}_{1} - \hat{x}(\tau_1) \right),  
    \end{equation}
    where $G(0,\tau_1)$ is the weighted controllability Gramian:
    \begin{equation}\label{eq:gt}
        G(0, \tau_1)=\int_{0}^{\tau_1} \e^{A(\tau_1-s)} B R^{-1} B^T \e^{A^T(\tau_1-s)} \dd s,
    \end{equation}
    where $R > 0$ is a control-weighting matrix as defined in \eqref{eq:cost_kd}. The term $\hat{x}(\tau_1)$ represents the state evolution at time $\tau_1$ in the absence of control input, given by:
    $$\hat{x}(\tau_1) = \e^{A(\tau_1-0)} \mathsf{x}_{0} + \int_{0}^{\tau_1} \e^{A (t-s)}d \ \dd s.$$
\end{theorem}
The proof of Theorem \ref{Thm1} is provided in Appendix \ref{appendix_proofs}.
\begin{remark}
    As the weight $\mu_1$ associated with the control cost term increases, the control strategy is adjusted to favor smoother and more gradual changes in the state and control variables. This leads to a reduction in the overall control effort, as the system aims to reach the final state with minimal energy expenditure. 
\end{remark}


\subsection{Optimal Arrival Time}
To compute the optimal trajectory without a fixed arrival time $\tau_1$, we introduce the optimal arrival time $\tau^*_1$ into the computation. To determine $\tau^*_1$, we first evaluate the cost of the trajectory from $\mathsf{x}_{0}$ and $\mathsf{x}_{1}$ for a given arrival time $\tau_1$. The cost function is expressed as:
\begin{equation}
    c[\tau]=\mu_1(\tau_1 + (\mathsf{x}_{1}-x(\tau_1))^TG^{-1}(0, \tau_1)(\mathsf{x}_{1}-x(\tau_1))) + \mu_2 c_{\mathrm{TLTL}},
\end{equation}
where $\mu_1$ and $\mu_2$ are weighting factors as in \eqref{eq:cost_fn}, and $c_{\mathrm{TLTL}}$ represents the cost associated with satisfying the temporal logic specification, which does not depend on $\tau_1$. The optimal arrival time $\tau^*_1$ is found by minimizing the total cost:
\begin{align}
    \tau^*_1 &= \mathrm{arg}\min_{\tau_1>0} c[\tau_1], \nonumber \\
    &= \{ \tau_1 \mid \dot{c}[\tau_1]=\mu_1(1 - 2(A\mathsf{x}_{1}+d)^T e(\tau_1) \nonumber \\
    &\qquad \qquad \quad - e(\tau_1)^T B R^{-1} B^T e(\tau_1)) = 0\},
\end{align}
where $e(\tau_1)=G(0,\tau_1)^{-1}(\mathsf{x}_{1} - \hat{x}(\tau_1))$, and $G(0,\tau_1)$ is the controllability Gramian. The optimal arrival time $\tau^*_1$ is the value of $\tau_1$ that sets the derivative of the cost $c[\tau_1]$ to zero, minimizing the cost and yielding the optimal trajectory.

\subsection{Optimal Trajectory between Two States}
Once the optimal arrival time $\tau^*_1$ is determined, the optimal trajectory $\pi^*[\mathsf{x}_{0}, \mathsf{x}_{1}]$ can be computed. From Equation \eqref{eq:ut}, the optimal control policy is given by:
\begin{equation} \label{eqn:con}
    u(t)=R^{-1} B^T y(t), \ y(t) = \e^{A^T(\tau^*_1-t)}e(\tau^*_1).
\end{equation}
Note, $\dot{y}(t) = -A^T y(t)$ and $y(\tau^*_1) = e(\tau^*_1)$.
The state trajectory $x(t)$ is then governed by the differential equation:
\begin{align}
    \dot{x}(t)=A x(t) + B R^{-1} B^T y(t) + d, \quad x(\tau^*_1)=\mathsf{x}_{1}.
\end{align}
This forms a coupled system of differential equations for $x(t)$ and $y(t)$. The joint dynamics is expressed as:
\begin{align}
\begin{bmatrix}
\dot{x}(t)  \\
\dot{y}(t) 
\end{bmatrix}
&=
\underbrace{\begin{bmatrix}
A & BR^{-1}B^T \\
0 & -A^T 
\end{bmatrix}}_{\hat A}
\begin{bmatrix}
x(t)  \\
y(t) 
\end{bmatrix}
+
\begin{bmatrix}
d  \\
0 
\end{bmatrix},\\
\begin{bmatrix}
x(\tau^*_1)  \\
y(\tau^*_1) 
\end{bmatrix}
&=
\begin{bmatrix}
\mathsf{x}_{1} \\
e(\tau^*_1) 
\end{bmatrix}.
\end{align}
The solution to this system is given by:
\begin{align} \label{eqn:traj}
\begin{bmatrix}
x(t) \\
y(t)
\end{bmatrix}
=
\e^{\hat A
(t - \tau^*_1)}
\begin{bmatrix}
\mathsf{x}_{1} \\
e(\tau^*_1)
\end{bmatrix}+
\int_{\tau^*_1}^{t}
\e^{\hat A (t - s)}
\begin{bmatrix}
d \\
0
\end{bmatrix}
\; \dd s.
\end{align}


\subsection{Extending to Overall Optimal Trajectory}
The problem of connecting two states optimally can be extended to compute the overall optimal trajectory between multiple states by leveraging the optimal substructure property of the cost function. This property ensures that the optimal solution for the entire trajectory can be constructed from optimal solutions of its subproblems, i.e., the segments of the trajectory between consecutive waypoints.

\begin{theorem}\label{thm2}
    The cost function defined in \eqref{eq:cost_fn} adheres to the optimal substructure property, i.e., for any intermediate time $0<t<\tau_1$, the total cost from $\mathsf{x}_{0}$ to $\mathsf{x}_{1}$ can be split into two parts:
    $$c^*[\mathsf{x}_{0}, \mathsf{x}_{1}] = c^*[\mathsf{x}_{0}, x(t)] + c^*[x(t), \mathsf{x}_{1}],$$
    where $x(t)$ represents the intermediate state at time $t$ along the optimal trajectory.
\end{theorem}

The proof of Theorem \ref{thm2} is provided in Appendix \ref{appendix_proofs}.

Therefore, an optimal collision-free trajectory $\pi_{\mathrm{free}}$ between a start state $\mathsf{x}_{\mathrm{start}}$ and a goal state $\mathsf{x}_{\mathrm{goal}}$ can be constructed by concatenating a sequence of optimal trajectories between successive intermediate states $\left( \mathsf{x}_{\text{start}}, \mathsf{x}_1, \mathsf{x}_2, \ldots, \mathsf{x}_{\text{goal}} \right)$ in the free space $X_{\text{free}}$. This ensures that the overall trajectory minimizes the total cost.

The algorithm to compute a kinodynamically feasible and optimal trajectory that adheres to temporal logic specification is presented below. 

\begin{algorithm}
\caption{}
\begin{algorithmic}[1]
    \State Initialize: $T \gets \{\mathsf{x}_\mathrm{start}\}$, $x \gets x_{\text{start}}$, $\pi^* \gets \{\}$, $u^* \gets \{\}$
    \While{True}
        \State Randomly sample $x_i \in X_\mathrm{free}$
        \State Linearize the system dynamics about $(\hat{x}, \hat{u})$ \eqref{eq:sysdynlinear}
        \State Calculate $c^*[\pi = \{x,x_i\},\psi]$ \eqref{eq:cost_fn}
        \State $x \gets arg \min_{\{(x \in T) \wedge \|x-x_i\| < r\}} \left( c[x, \psi] + c^*[\{x,x_i\}, \psi] \right)$
        \State $c[x_i, \psi] \gets c[x, \psi] + c^*[\{x,x_i\}, \psi]$
        \State $\text{parent}[x_i] \gets x$
        \State $\pi^* = \pi^* \cup \text{tr}\{x, x_i\}$ \eqref{eqn:traj}, \ $u^* = u^* \cup \text{con}\{x, x_i\}$ \eqref{eqn:con}
        \For{\{$x \in T \cup \{\mathsf{x}_\mathrm{goal}\} \mid \|x-x_i\| < r$ \} }
            \If{$c[x, \psi] > c[x_i, \psi] + c^*[\{x_i, x\}, \psi]$}
                \State $c[x, \psi] \gets c[x_i, \psi] + c^*[\{x_i, x\}, \psi]$
                \State $\text{parent}[x] \gets x_i$
                \State $\pi^* = \pi^* \cup \text{tr}\{x_i, x\}$ \eqref{eqn:traj}, 
                \State $u^* = u^* \cup \text{con}\{x_i, x\}$ \eqref{eqn:con}
            \EndIf
        \EndFor
        \State $T \gets T \cup \{x_i\}$
    \EndWhile
    \State \Return $(\pi^*, u^*)$
\end{algorithmic}
\end{algorithm}

\section{Experimental Results}
In this section, we present the experimental results for two different dynamical systems: 1) an omnidirectional mobile robot and 2) a quadrotor, each tested with various complex TLTL specifications.
In the following case studies, each region, $M$, is modeled as a hyperrectangle with center $\mathrm{center}(M)$ and dimensions, $d_{M}$. We define the cooresponding predicate function as $\tilde{M} := \|x - \mathrm{centre}(M)\|_{\infty} < d_{M}$. 

\begin{figure}[H]
    \centering
     \begin{subfigure}[b]{0.23\textwidth}
         \centering
         \includegraphics[width=\textwidth]{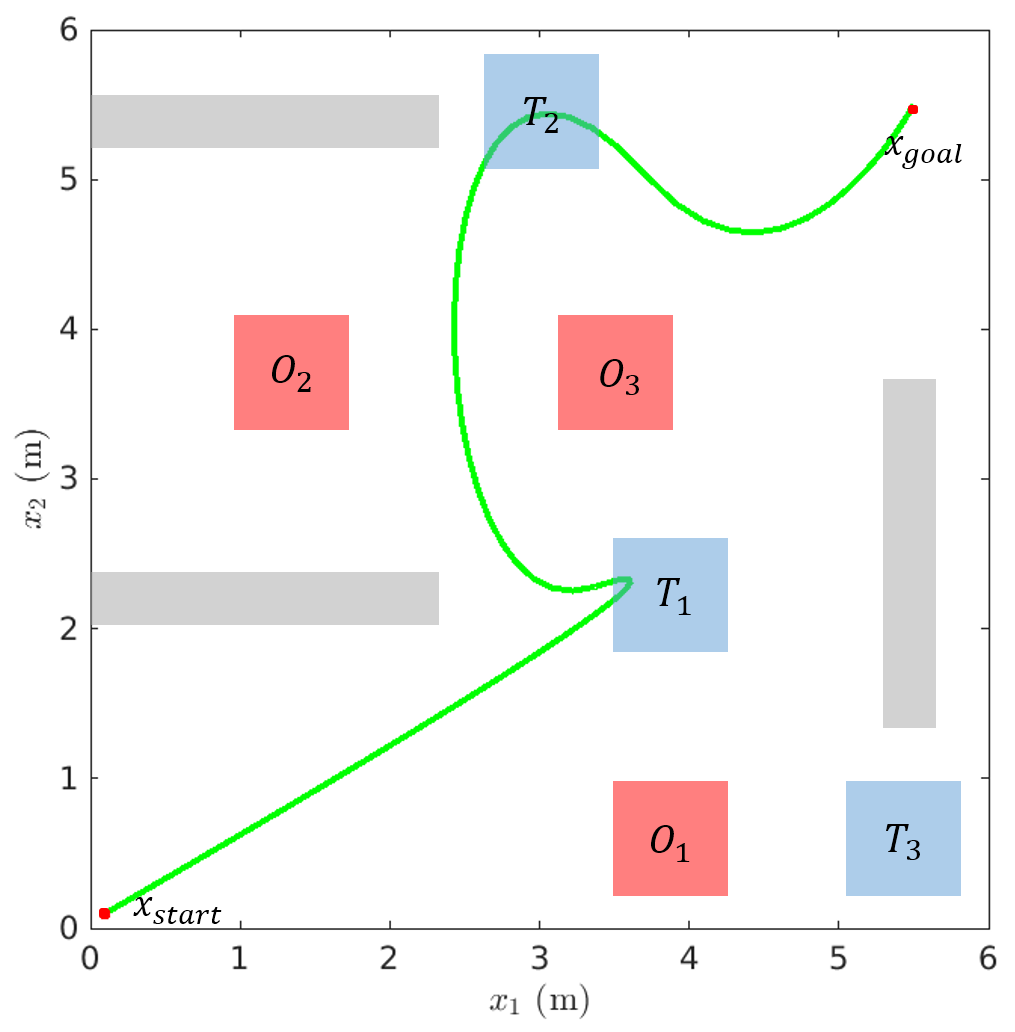}
         \caption{Robot chooses path 1}
         \label{fig:or1}
     \end{subfigure}
     \hfill
     \begin{subfigure}[b]{0.23\textwidth}
         \centering
         \includegraphics[width=\textwidth]{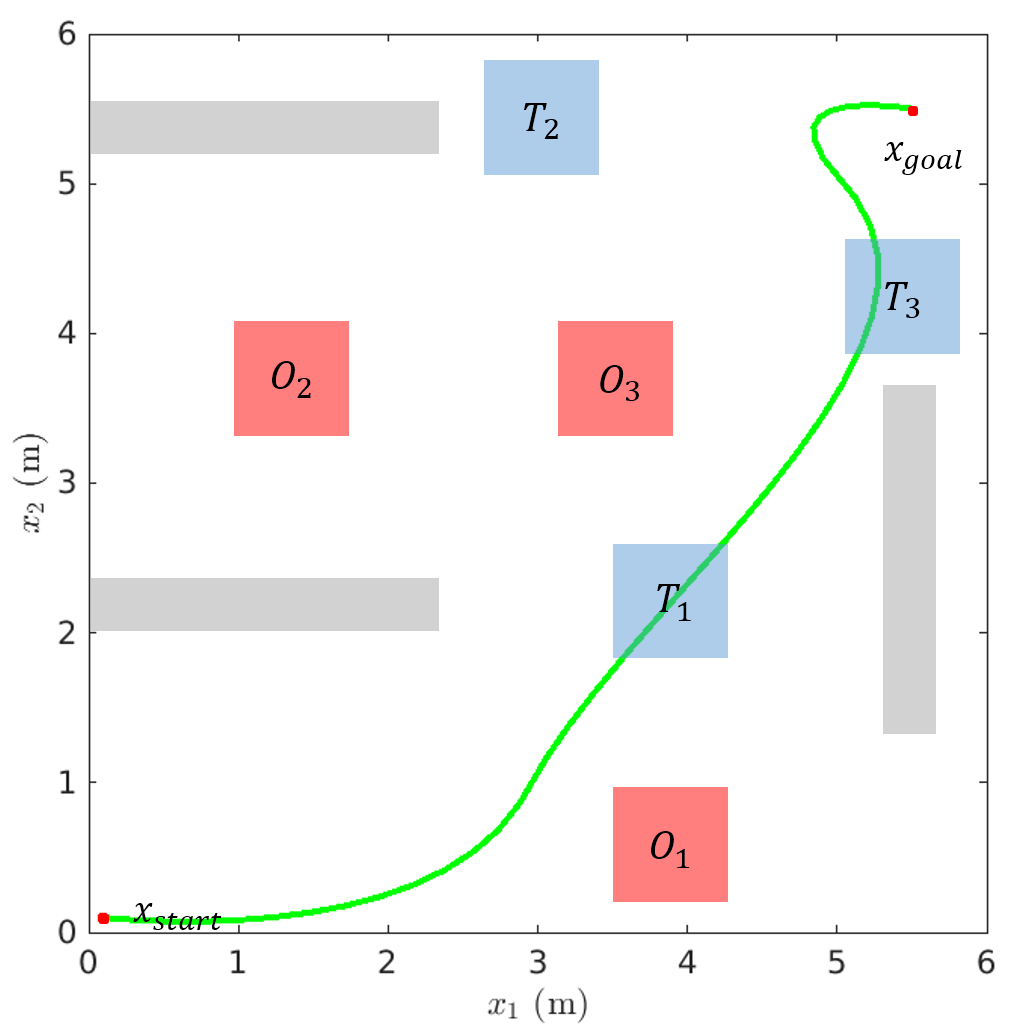}
         \caption{Robot chooses path 2}
         \label{fig:or2}
     \end{subfigure}
    \caption{Simulation results for mobile robot under Specification $\psi_1$.}
    \label{fig:psi1}
\end{figure}

\subsection{Mobile Robot Navigation}
This case study focuses on a mobile robot operating in a 2D environment $X \subset \mathbb{R}^2$. The vehicle dynamics follow the model presented in \cite{STT}. The workspace is defined as a square region measuring $X = [0, 6] \ m \times [0, 6] \ m$ and contains several obstacles, which are marked in grey. We impose velocity constraints on the robot to ensure safe and efficient operation, limiting its speed to $[-0.22, 0.22] \ m/s$. The control penalty is set to $R = I$. With these constraints and dynamics in place, we aim to derive the optimal trajectory for the robot based on two different temporal logic specifications. 

\subsubsection{Specification 1}
The initial state is $\mathsf{x}_{\mathrm{start}} = [0.1,0.1] \ m$ and the target set is $\mathsf{x}_{\mathrm{goal}} = [5.5,5.5] \ m$. The vehicle's task is to \textit{"eventually"} visit region $T_1$, and \textit{"eventually"} visit either region $T_2$ or $T_3$, and to \textit{"always"} avoid the regions $O_1, O_2 $ and $O_3$. The LTL specification for the task can be expressed as follows:
$$\psi_1 = \Diamond \tilde T_1 \wedge \Diamond ( \tilde T_2 \vee \tilde T_3) \wedge \Box (\neg \tilde O_1 \wedge \neg \tilde O_2 \wedge \neg \tilde O_3).$$

\begin{figure}[H]
    \centering
     \begin{subfigure}[b]{0.23\textwidth}
         \centering
         \includegraphics[width=\textwidth]{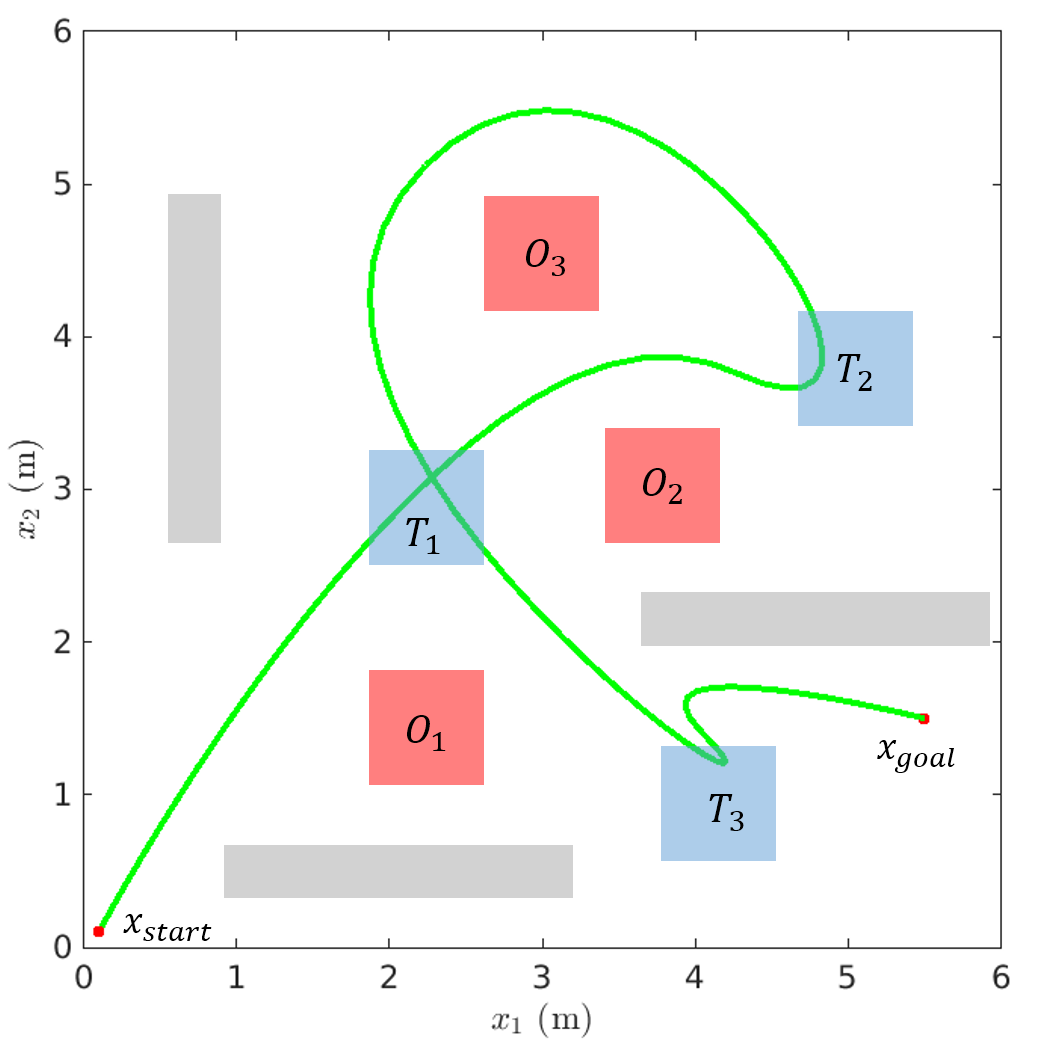}
         \caption{With KD constraints}
         \label{fig:withkd}
     \end{subfigure}
     \hfill
     \begin{subfigure}[b]{0.23\textwidth}
         \centering
         \includegraphics[width=\textwidth]{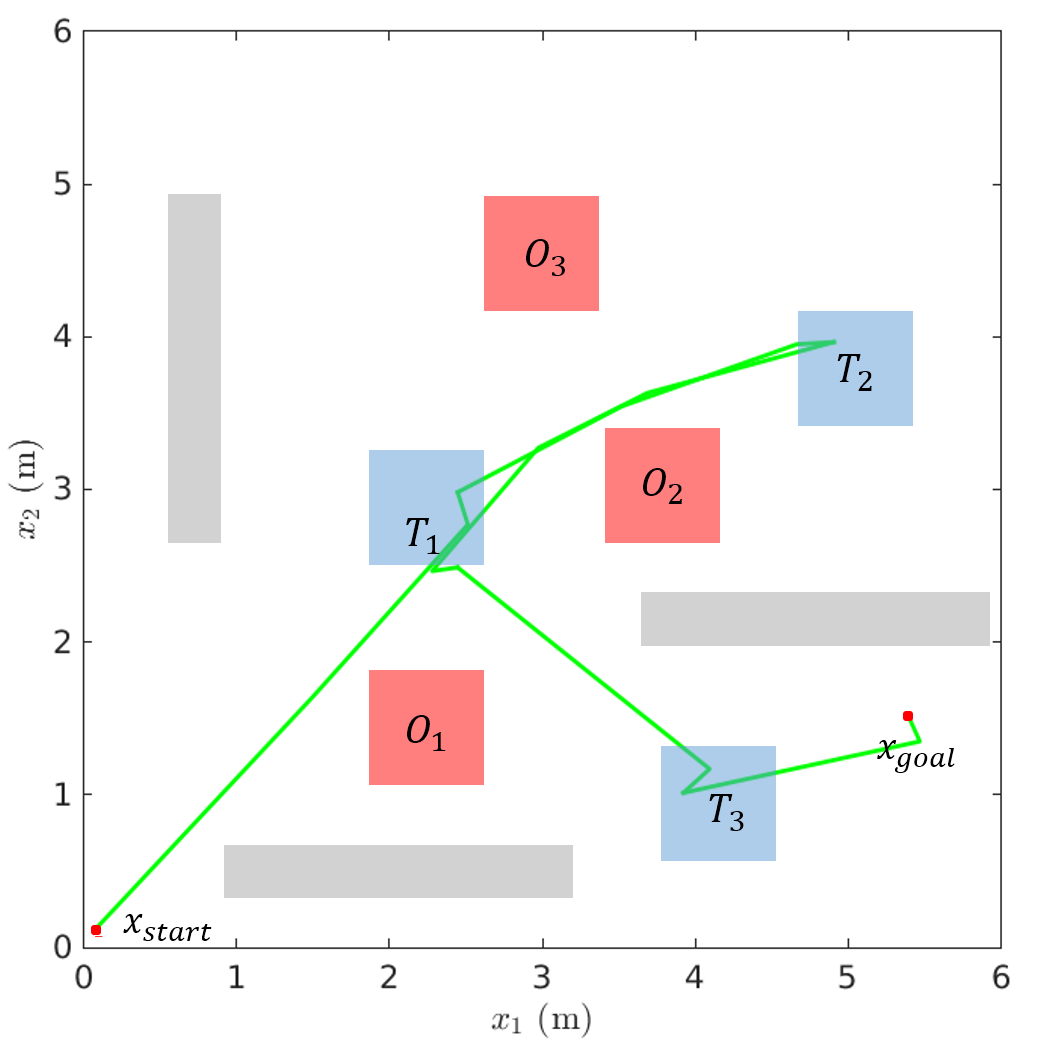}
         \caption{Without KD constraints}
         \label{fig:withoutkd}
     \end{subfigure}
    \caption{Comparison of simulation results with and without kinodynamic (KD) constraints for mobile robot under Specification $\psi_2$.}
    \label{fig:psi2}
\end{figure}

It is important to highlight that the proposed algorithm is designed to always return the optimal trajectory based on the given temporal logic specifications. When the robot is presented with the option to "eventually" visit either region $T_2$ or $T_3$, the algorithm selects the path that minimizes the overall cost, which typically corresponds to the shortest or most efficient route.

For instance, in the scenario in Figure \ref{fig:or1}, the robot consistently chooses to pass through region $T_2$ because it offers a shorter path to the goal. However, if we change the environment by shifting the location of $T_3$, as illustrated in Figure \ref{fig:or2}, the algorithm adapts accordingly. It now identifies $T_3$ as the more optimal region to visit, since it provides a shorter route under the new conditions. This adaptability demonstrates the algorithm's capability to dynamically find the most efficient trajectory in response to changes in the workspace

\subsubsection{Specification 2}
The initial state is $\mathsf{x}_{\mathrm{start}} = [0.1,0.1]\ m$ and the target set is $\mathsf{x}_{\mathrm{goal}} = [5.5,1.5] \ m$. The vehicle's task is to \textit{"always eventually"} visit region $T_1$, and \textit{"eventually"} visit regions $T_2$ and $T_3$, and to \textit{"always"} avoid regions $O_1, O_2 $ and $O_3$. The LTL specification for this task is given by:
$$\psi_2 = \Box ( \Diamond \tilde T_1) \wedge \Diamond \tilde T_2 \wedge \Diamond \tilde T_3 \wedge \Box (\neg \tilde O_1 \wedge \neg \tilde O_2 \wedge \neg \tilde O_3).$$

The videos showcasing real-world demonstrations, illustrated in Figure \ref{fig:rw}, are available at \href{https://indianinstituteofscience-my.sharepoint.com/:f:/g/personal/ratnangshud_iisc_ac_in/EiOAlrGImJRGgn60IfaNlXkB0D5khmFqaYb1bncYEDQC8A?e=XaDM8R}{[Link]}. These demonstrations highlight the practical implementation and effectiveness of the proposed approach in obstacle-laden environments.


\subsection{Navigation of UAV in 3D Space}
\begin{figure}[H]
\centering
\includegraphics[width=0.3\textwidth]{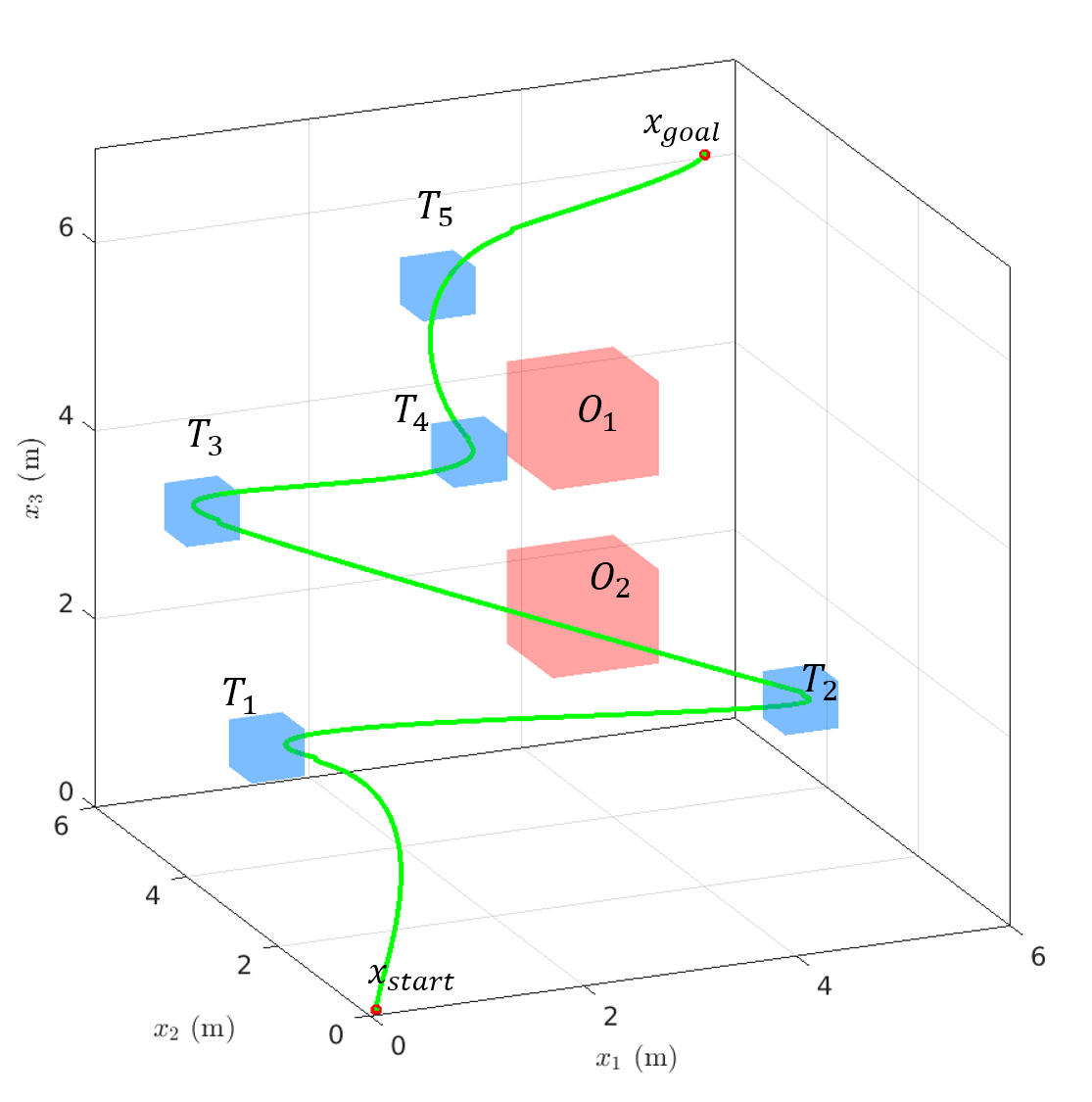}
\caption{Navigation of UAV under Specification $\psi_3$.}
\label{fig:3d}
\end{figure}

In this case study, we examine a UAV operating under double-integrator dynamics, which have been adapted from \cite{APF}. The UAV navigates within an obstacle-laden three-dimensional environment $X \subset \mathbb{R}^3$ with spatial boundaries defined as $[0, 6]\ m \times [0, 6]\ m \times [0, 6]\ m$. Obstacles within this space are represented in gray. We limit the maximum thrust on the drone to $[-5, 5] N$
The initial state is $\mathsf{x}_{\mathrm{start}} = [0.1,0.1,0.1] \ m$ and the target set is $\mathsf{x}_{\mathrm{goal}} = [5.5, 5.5, 6.25] \ m$. The vehicle's task is to \textit{"eventually"} visit region $T_1$, $T_2, T_3, T_4$ and $T_5$, and to \textit{"always"} avoid the region $O$. The LTL specification is:
$$\psi_3 = \Diamond \tilde T_1 \wedge \Diamond \tilde T_2 \wedge \Diamond \tilde T_3 \wedge \Diamond \tilde T_4 \wedge \Diamond \tilde T_5 \wedge \Box (\neg \tilde O).$$
The simulation results are shown in Figure \ref{fig:3d}.

\section{Comparative Analysis}
In this section, we compare the performance of the proposed algorithm with symbolic control techniques \cite{tabuada2009verification}, and LTL path planning algorithms that do not consider kinodynamic constraints \cite{linard2023real}.

Firstly, while symbolic control techniques can address input constraints, they are computationally intensive. Sampling-based approaches and traditional LTL path planning algorithms \cite{smith2010optimal} ignore kinodynamic constraints and result in jagged trajectories with sharp turns, as illustrated in Figure \ref{fig:withoutkd}. Following these jagged paths with standard control techniques, such as PID \cite{normey2001mobile}, leads to control input spikes that frequently exceed the system’s operational limits.

\begin{table}[H]
\centering
\caption{Comparison with standard techniques}
\resizebox{.48\textwidth}{!}{
\begin{tabular}{c|ccc|ccc|}
\cline{2-7}
\multicolumn{1}{l|}{\multirow{2}{*}{}} & \multicolumn{3}{c|}{Computation Time} & \multicolumn{3}{c|}{Max. Control Effort} \\ \cline{2-7} 
\multicolumn{1}{l|}{} & \multicolumn{1}{c|}{$\psi_1$} & \multicolumn{1}{c|}{$\psi_2$} & $\psi_3$ & \multicolumn{1}{c|}{$\psi_1$} & \multicolumn{1}{c|}{$\psi_2$} & $\psi_3$ \\ \hline
\multicolumn{1}{|c|}{\textbf{\begin{tabular}[c]{@{}c@{}}Proposed \\ approach\end{tabular}}} & \multicolumn{1}{c|}{\textbf{87 s}} & \multicolumn{1}{c|}{\textbf{116 s}} & \textbf{795 s} & \multicolumn{1}{c|}{\textbf{0.199 m/s}} & \multicolumn{1}{c|}{\textbf{0.204 m/s}} & \textbf{1.056 N} \\ \hline
\multicolumn{1}{|c|}{\begin{tabular}[c]{@{}c@{}}Symbolic\\ Control\end{tabular}} & \multicolumn{1}{c|}{626 s} & \multicolumn{1}{c|}{861 s} & 35146 s & \multicolumn{1}{c|}{0.22 m/s} & \multicolumn{1}{c|}{0.22 m/s} & 5 N \\ \hline
\multicolumn{1}{|c|}{\begin{tabular}[c]{@{}c@{}}w/o KD \\ contraints\end{tabular}} & \multicolumn{1}{c|}{75 s} & \multicolumn{1}{c|}{99 s} & 714 s & \multicolumn{1}{c|}{1.386 m/s} & \multicolumn{1}{c|}{4.838 m/s} & 22.395 N \\ \hline
\end{tabular}
}
\label{tab:1}
\end{table}

We summarize the comparison results in Table \ref{tab:1}. These results demonstrate the proposed algorithm's superiority in both computational efficiency and in generating kinodynamically feasible paths with reduced control effort that directly contributes to more energy-efficient and reliable operation.





\section{CONCLUSIONS}
In this paper, we addressed the gap between temporal logic-based motion planning and kinodynamic feasibility by introducing an RRT*-based method that incorporates both aspects. By defining a cost function that balances temporal logic robustness with kinodynamic constraints, our approach ensures that generated trajectories are not only logically correct but also physically implementable. This integration allows for planning in constraint-heavy environments, paving the way for more versatile applications in robotics and autonomous systems. Our simulations and real-world experiments validate the approach, showing its potential to generate optimal paths that meet temporal logic requirements while respecting the system's motion capabilities. Future work may involve extending this framework to handle more complex dynamic and stochastic environments, and real-world multi-agent scenarios.

\begin{appendices}
\section{}\label{appendix_proofs}

\subsection*{Proof of Theorem \ref{Thm1}}
\label{proof:thm1}

\begin{proof}
    To derive the optimal control law for the linear system with a fixed final state $x(\tau_1) = \mathsf{x}_{1}$, we consider a quadratic cost function of the form:
    $$J = \int_{0}^{\tau_1} \left( \mu_1 + \mu_1\frac{1}{2} u(t)^T R u(t) - \mu_2 \left( \frac{\partial \hat{\rho}^{\psi}}{\partial x} \right)^T\dot{x} \right) \dd t,$$
    where $R > 0$ is a control-weighting matrix. 

    We first form the Hamiltonian for this problem:
    \begin{align*}
        H(x,u,\lambda,t) &= \mu_1\frac{1}{2} u(t)^T R u(t) \ + \nonumber \\
        &\left( \lambda(t) - \mu_2\frac{\partial \hat{\rho}^{\psi}}{\partial x} \right)^T(Ax(t) + Bu(t) + d),    
    \end{align*}
    where $\lambda(t)$ is the costate (Lagrange multiplier) vector, which enforces the system dynamics.

    To find the optimal control $u(t)$, we apply Pontryagin’s Minimum Principle. This gives us
    the stationary condition for optimal control:
    \begin{align}\label{eqp:oc1}
        &\frac{\partial H}{\partial u} = 0 \implies \mu_1 R u(t) + B^T \left(\lambda(t) - \mu_2\frac{\partial \hat{\rho}^{\psi}}{\partial x} \right) = 0 \nonumber \\
        &\implies u(t) = - \frac{1}{\mu_1} R^{-1} B^T \left(\lambda(t) - \mu_2\frac{\partial \hat{\rho}^{\psi}}{\partial x} \right),
    \end{align}
    and the costate dynamics: 
    \begin{align}\label{eqp:cos}
        &\dot{\lambda}(t) = -\frac{\partial H}{\partial x} = -A^T \left(\lambda(t) - \mu_2\frac{\partial \hat{\rho}^{\psi}}{\partial x} \right).
    \end{align}
    Let $\theta(t) := \lambda(t) - \mu_2\frac{\partial \hat{\rho}^{\psi}}{\partial x}$. Then, we have
    \begin{align} \label{eqp:th}
        u(t) &= - \frac{1}{\mu_1} R^{-1} B^T \theta(t), \\   
        \dot{\theta}(t) &= -A^T\theta(t) \implies \theta(t) = \e^{A^T (\tau_1 - t)} \theta(\tau_1),
    \end{align}
    where $\theta(\tau_1)$ is still unknown. 
    Substituting the above in the system dynamics \eqref{eqn:sysdyn} yields
    \begin{align}\label{eqp:x}
        &\dot{x} = Ax - B R^{-1} B^T \e^{A^T (\tau_1 - t)} \theta(\tau_1) + d \nonumber \\
        \implies &x(t) = \e^{A(t - 0)}\mathsf{x}_{0} + \int_{0}^{t} \e^{A (t-s)}d \ \dd s \nonumber \\
        & - \int_{0}^{t} \e^{A(t-s)} B R^{-1} B^T \e^{A^T (\tau_1 - s)} \theta(\tau_1) \dd s.
    \end{align}
    Now, to find $\theta(\tau_1)$, evaluate the value of $x(t)$ at $t=\tau_1$,
    \begin{equation}\label{eqp:x2}
        x(\tau_1) = \mathsf{x}_{1} = \hat{x}(\tau_1) - G(0, \tau_1) \theta(\tau_1)    
    \end{equation}
    where $\hat{x}(\tau_1) = \e^{A(\tau_1 - 0)}\mathsf{x}_{0} + \int_{0}^{\tau_1} \e^{A (t-s)}d \ \dd s$ is the state response at time $t = \tau_1$ in the absence of control input, and $G(0, \tau_1)=\int_{0}^{\tau_1} \e^{A(\tau_1-s)} B R^{-1} B^T \e^{A^T(\tau_1-s)} \dd s$ is the weighted controllability Gramian.
    Therefore, rearranging \eqref{eqp:x2}, we get,
    \begin{equation}\label{eqp:lam}
        \theta(\tau_1) = -G^{-1}(0, \tau_1) \left( \mathsf{x}_{1} - \hat{x}(\tau_1) \right).
    \end{equation}
    Finally, the optimal control can be written by substituting \eqref{eqp:lam} in \eqref{eqp:th},
    \begin{equation}\label{eqp:oc}
      u(t) = \frac{1}{\mu_1} R^{-1} B^T \e^{A^T(\tau_1-t)} G^{-1}(0, \tau_1) \left(\mathsf{x}_{1} - \hat{x}(\tau_1)\right).  
    \end{equation}
\end{proof}

\subsection*{Proof of Theorem \ref{thm2}}
\label{proof:thm2}

\begin{proof}
    Let $\pi^*[\mathsf{x}_{0}, \mathsf{x}_{1}] = (x[\cdot], u[\cdot], \tau_1)$ represent the optimal trajectory between two states $\mathsf{x}_{0} \in X$ and $\mathsf{x}_{1} \in X$. The cost associated with this optimal trajectory is denoted by $c^*[\mathsf{x}_{0}, \mathsf{x}_{1}]$, and it is defined as:
    \begin{align*}
        &c^*[\mathsf{x}_{0},\mathsf{x}_{1}] = \mu_1 c_{C}[\pi^*] + c_{TLTL}[\pi^*, \psi], \\
        &= \mu_1 \int_{0}^{\tau_1} \left(1 + \frac{1}{2} u^*(s)^T R u^*(s) \right) + \mu_2 \int_{0}^{\tau_1} -\left( \frac{\partial \hat{\rho}^{\psi}}{\partial x} \right)^T \dot{x} \ \dd s, \\
        &= \int_{0}^{t} \mu_1 \left(1 + \frac{1}{2} u^*(s)^T R u^*(s) \right) - \mu_2 \left( \frac{\partial \hat{\rho}^{\psi}}{\partial x} \right)^T \dot{x} \ \dd s \\
        &\ \ \ + \int_{t}^{\tau_1} \mu_1 \left(1 + \frac{1}{2} u^*(s)^T R u^*(s) \right) - \mu_2 \left( \frac{\partial \hat{\rho}^{\psi}}{\partial x} \right)^T \dot{x} \ \dd s, \\
        &= c^*[\mathsf{x}_{0},x(t)] + c^*[x(t),\mathsf{x}_{1}].
    \end{align*}
\end{proof}
\end{appendices}
\addtolength{\textheight}{-12cm}

\bibliography{sources}

\end{document}